# Heat Reduction in Erbium Doped Microresonator

**Nitzan Shani[1], Bar Favelukis[1], Menachem Guttman[1], Pavel Sidorenko[2], Raz Gvishi[1], Tal Carmon[1]**

[1] *School of Electrical Engineering, Tel Aviv University, Tel Aviv 6997801, Israel*

[2] *Andrew and Erna Viterbi Faculty of Electrical and Computer Engineering, Technion – Israel Institute of Technology, Haifa 3200003, Israel*

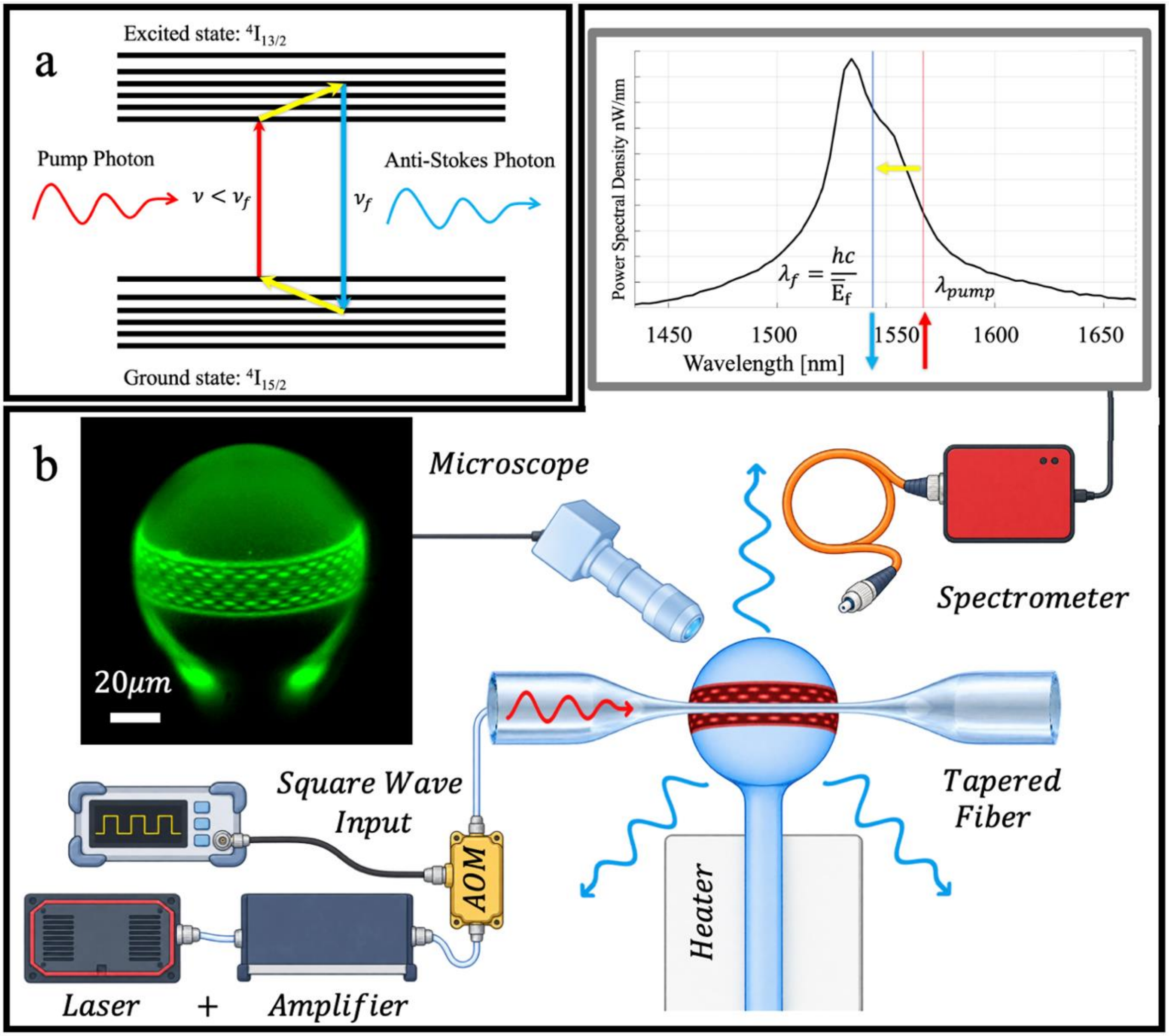

**Figure 1. Experimental setup**. **a**. is based on Bahae and epstein[1] describing the transition between the ground state $^4I_{15/2}$ to the excited state $^4I_{13/2}$. The many lines represent the Stark sublevels within the energy level. The pump (red) transitioning (yellow) into fluorescent photon (blue) with higher energy. **b**. In the setup, an AOM is introduces amplified 1560nm square wave signal to the resonator. The signal going through 3 photon upconversion emits green photons (left). The resonator dipped in $Er^{+3}$ sol-gel was placed on a TEC board varying the resonator's temperature from 25 up to 62ºC. The spectrometer measures the power spectral density. The black spectrum is the average of 3 different repetitions; each one is of 1000 spectrums measured consequently. On the spectrum, the same process as in (a) is shown for pump(red), transition(yellow) and average emitted energy(blue).

## 1. Introduction

Anti-Stokes fluorescence can remove thermal energy from an optically excited material when the average fluorescence photon caries more energy than the absorbed pump photon. Following the pioneering demonstration of net cooling in a solid by Epstein et al. [1], the principles and implementations of optical refrigeration were later discussed by Epstein and Sheik-Bahae [2]. Since then, fluorescence cooling and heat-reduction were discussed and demonstrated in rare-earth-doped optical fibers, also serving as radiation balanced amplifiers [3,4], cavity-based systems [5,6]. Laser refrigeration by fluorescence has also been investigated in nanoparticle-coated whispering-gallery-mode (WGM) microspheres [7].

In cavity, pump absorption is enhanced by repeatedly passing through the cooling medium. Early implementations included Brewster-cut ytterbium-doped ZBLAN glass inside a laser resonator and coated ZBLAN samples in two-mirror enhancement cavities [1,6,8]. Their optical losses consist of mirror losses, scattering, and residual interface reflection, including departures from ideal Brewster-angle operation. WGM microresonators provide an alternative in which circulating light is confined by total internal reflection at the dielectric boundary.

A cooled resonator will absorb heat from its warmer surroundings through the surface of the microresonator. There is a tradeoff here when discussing micro-sized resonators. Shrinking the resonator reduces both its surface area and its volume, but the volume decreases faster since surface area scales as [$m^2$] and volume as [$m^3$]. Once the resonator is in net cooling state, smaller devices can consequently face a larger environmental heat load per unit volume.

In this paper, we assess the potential for fluorescence cooling in silica WGM microresonators approximately 100 μm in diameter, coated with $Er^{+3}$-doped sol–gel. We examine the emitted spectrum and its dependence on pump wavelength and externally controlled temperature. Presenting higher-order modes, characterized by radial, azimuthal, and polar (axial) mode numbers, to cover a larger volume and by that causing heat reduction in the resonator. Going beyond a narrow equatorial ring. In particular, higher radial and polar orders offer different spatial excitation profiles. Their potential cooling benefit must be evaluated through their overlap with the doped coating and their optical losses.

## 2. Experimental setup and methods

We examine the fluorescence of an $Er^{3+}$-doped sol–gel coating on a whispering-gallery microresonator while optically exciting the device near 1.55 μm. Figure 1(a) illustrates the relevant $^4I_{15/2}$ and $^4I_{13/2}$ manifolds. Excitation from thermally populated Stark sublevels can be followed by fluorescence whose mean photon energy exceeds that of the pump. Figure 1(b) shows the optical path, an image of 3 photon upconversion from the resonator, and the measured NIR fluorescence spectrum.

### 2.1 Er-doped microsphere preparation

The Er-doped sol–gel film on the 106-μm-diameter silica sphere was prepared in a similar way to the method used by Yang et al. [3], with some changes. The sol–gel solution was prepared by hydrolyzing tetraethoxysilane (TEOS) in ethanol using water, with a TEOS:water:ethanol molar ratio of 1:2:4. Hydrochloric acid (HCl) was used as the acid catalyst for the sol–gel reaction. $Er^{3+}$ ions were introduced by adding erbium(III) acetate hydrate, previously dissolved in an HCl/water solution. For the resonator presented here, the weight ratio of erbium(III) acetate hydrate:water:32% (w/w) HCl was 8.27:7.0:4.7, resulting in an $Er^{3+}$ ion concentration of $1.0 \times 10^{21}$ ions $cm^{-3}$ in the solution. TEOS and ethanol were mixed for 1 h. The Er/HCl/water solution was then added dropwise, and the mixture was stirred for an additional 2 h. The solution was left to hydrolyze in a closed vial for at least 24 h before coating. The silica sphere on the optical fiber was immersed in the sol–gel solution for a few seconds and then withdrawn quickly. The coated sphere was left to dry at room temperature for about 2 days in low humidity nytrogen container. Final drying was performed with a two 0.1 seconds arc-plasma pulses, used to sinter the sol–gel coating to silica. Taking into account an estimated volume shrinkage factor of approximately 1.5 during drying, the $Er^{3+}$ concentration in the final sol–gel film was estimated to be $1.5 \times 10^{21}$ ions $cm^{-3}$.

### 2.2 Optical setup and temperature control

Before operation of the entire experimental system, the Quality Factor of the resonator was measured to be $Q = 2 \cdot 10^5$ at 1550nm. Light from an amplified tunable laser was modulated by an acousto-optic modulator (AOM) through the 1st order exit and coupled to the resonator through a tapered fiber. The optical power measured at the tapered fiber was approximately 20 mW. The AOM operated at 20 Hz with a duty cycle of 64%, turning the pump on and off at that rate. Light transmitted through the taper was put into a photodetector, while a microscope imaged the excited resonator modes. A multimode fiber with a $600\,\mu m$ core collected fluorescence and delivered it to a spectrometer covering $850 - 1700\,nm$. The spectrometer resolution was $5.46 \pm 0.86nm\ (mean \pm SD)$. The resonator was mounted on a temperature-controlled stage. We varied the stage set point between 25 and 62 °C and used the measured resonator's temperature in the analysis.

### 2.3 Spectral acquisition and background subtraction

For each measurement, we acquired three independent runs of 1000 consecutive spectra. The spectrum of the pump was measured separately and is indicated by the red vertical line in Fig. 1(b). The fluorescence spectra were acquired with a $200\,\mu sec$ delay after the AOM signal is completely off, allowing the pump inside the resonator to decay. Since the $Er^{+3}$ excited state lifetime is $10\,msec$, we had set the measurements integral to be $10\,msec$. Dark current was measured beforehand with 1000 spectrums as well.

In addition, the integration interval was bounded by the wavelengths at which the signal fell to 1% of its peak value.

### 2.4 Mean fluorescence photon energy

With the final spectrum form, we define the mean emitted photon energy as:

1) $\overline{E}_f = \frac{\int h\nu_f \cdot S_{emission}(\nu_f)}{\int S_{emission}(\nu_f)}$ ; $\nu_f = \frac{c}{\lambda_{meas}}$

The blue vertical line in Fig. 1(b) marks $\lambda_f$. The pump wavelength $\lambda_{pump}$ was obtained from independently measuring the pump spectrum.

## 3. Results

Figure 1(b) shows the fluorescence spectrum of the $Er^{3+}$-doped resonator together with the independently measured pump wavelength. The emission spectrum obtained shows tendancy to be higher to the left of the pump wavelength. The blue line marks the mean fluorescence wavelength defined in Eq. (1), and the red line marks the pump.

### 3.1 Mode-dependent fluorescence patterns

Figure 2 shows visible emission patterns obtained by exciting different resonances of the same $Er^{3+}$-coated microsphere. The spatial emission pattern changes with the excited resonance.

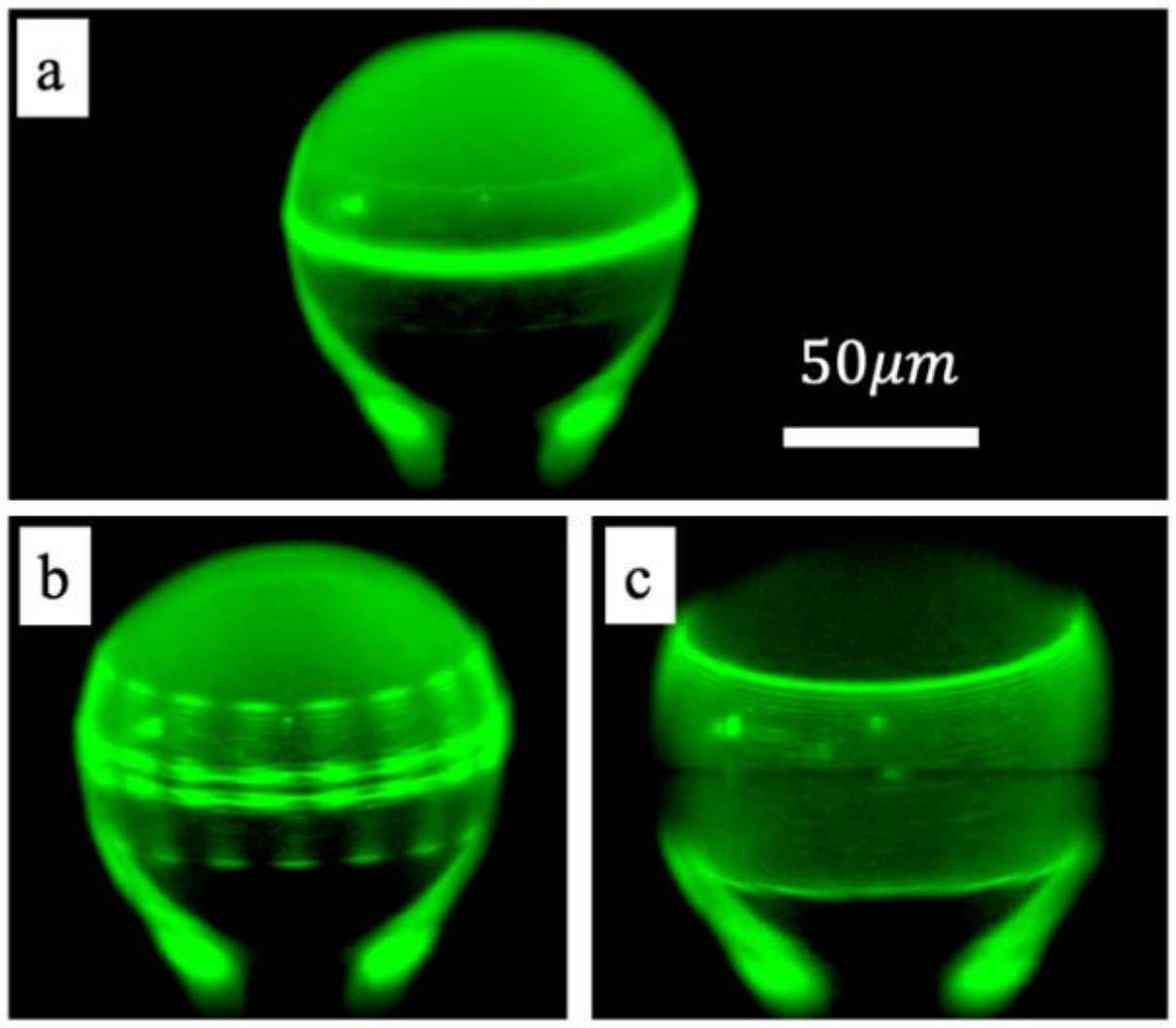

**Figure 2. Experimental results, fluorescent mode mapping: Spatial patterns of different modes**. The photos (a)-(c) were taken for different wavelengths between 1565 and 1570nm. The photos are an average combination of 5-7 photos of the same mode.

### 3.2 Fluorescence spectrum and energy analysis

We next compare the spectral contributions on either side of the pump for $\lambda_{pump} = 1568.05\,nm$. Figure 3(a) shows the fluorescence power spectral density, scaled by a factor of 25.25 based on Eq. 2.1. and the multimode fiber's NA:

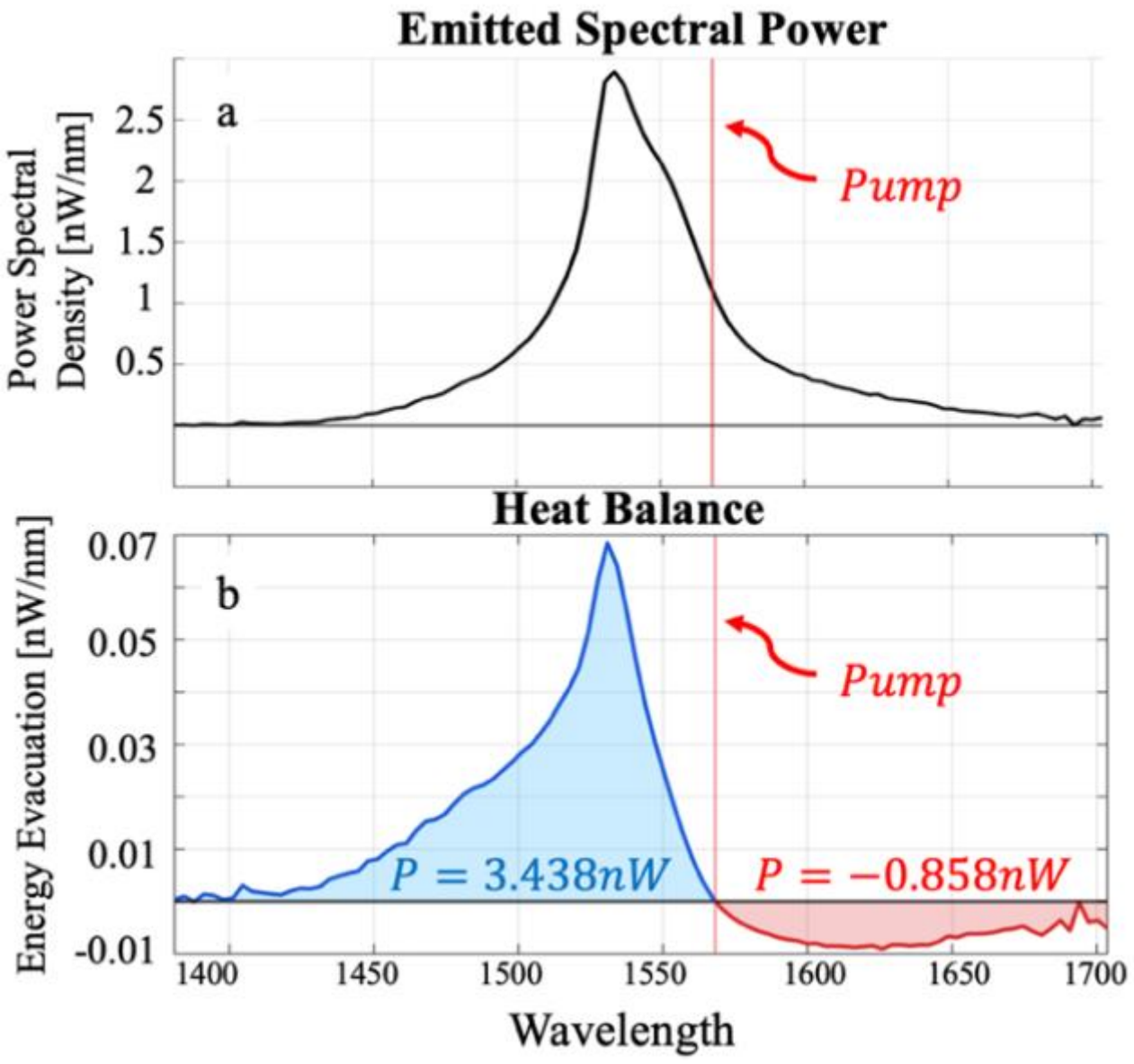

**Figure 3. Experimental results: Total power exiting the resonator visualization**. **a**. Total power exiting the resonator with red line as indication for pump at 1568.05nm. the original measured spectrum was multiplied by 25.25 since the multimode fiber numerical aperture covers 3.96% of fluorescence emitted from the resonator. **b.** Each frequency of the measured spectrum is compared to the pump frequency. When higher than pump frequency – more energy is leaving rather than introduced to the resonator (blue). When lower, less energy is leaving rather than coming in (red). At pump wavelength the incoming power equals the exiting power. Total power measured leaving the resonator is $2.58nW$

2)

2.1) $$S_{\lambda,total} = \frac{4\pi r^2}{\int_{\theta=0}^{NA}\int_{\varphi=0}^{360} r^2 sin\theta \, d\varphi d\theta} \cdot S_{\lambda,detected}$$

$$= 25.25 \cdot S_{\lambda,detected}$$

Figure 3(b) weights each part of the spectrum by its normalized $\Delta\nu$ difference from the pump:

2.2) $$25.25 \cdot \frac{\nu - \nu_p}{\nu_p}$$

Integration over Figure 3(b) will provide us with the power. The blue and red regions gives +3.438 nW and -0.858 nW respectively, with the summation of +2.580 nW. The dominance of the blue contribution shows that the detected fluorescence carries an excess photon-energy component.

### 3.3 Temperature dependence of fluorescence

To examine the temperature dependence of the fluorescence, we held the pump at 1565 nm and varied the temperature-controller set point over five values between 25 and 62 °C. Figure 4 shows that the mean fluorescence emitted photon energy increases with the substrate's temperature. A linear fit gives a slope of 8.23 $\mu eV\ °C^{-1}$ with ($R^2 = 0.9589$). The error bars show the standard deviation of the three repetitions at each temperature. This result describes the response of the fluorescence spectrum to externally applied heating.

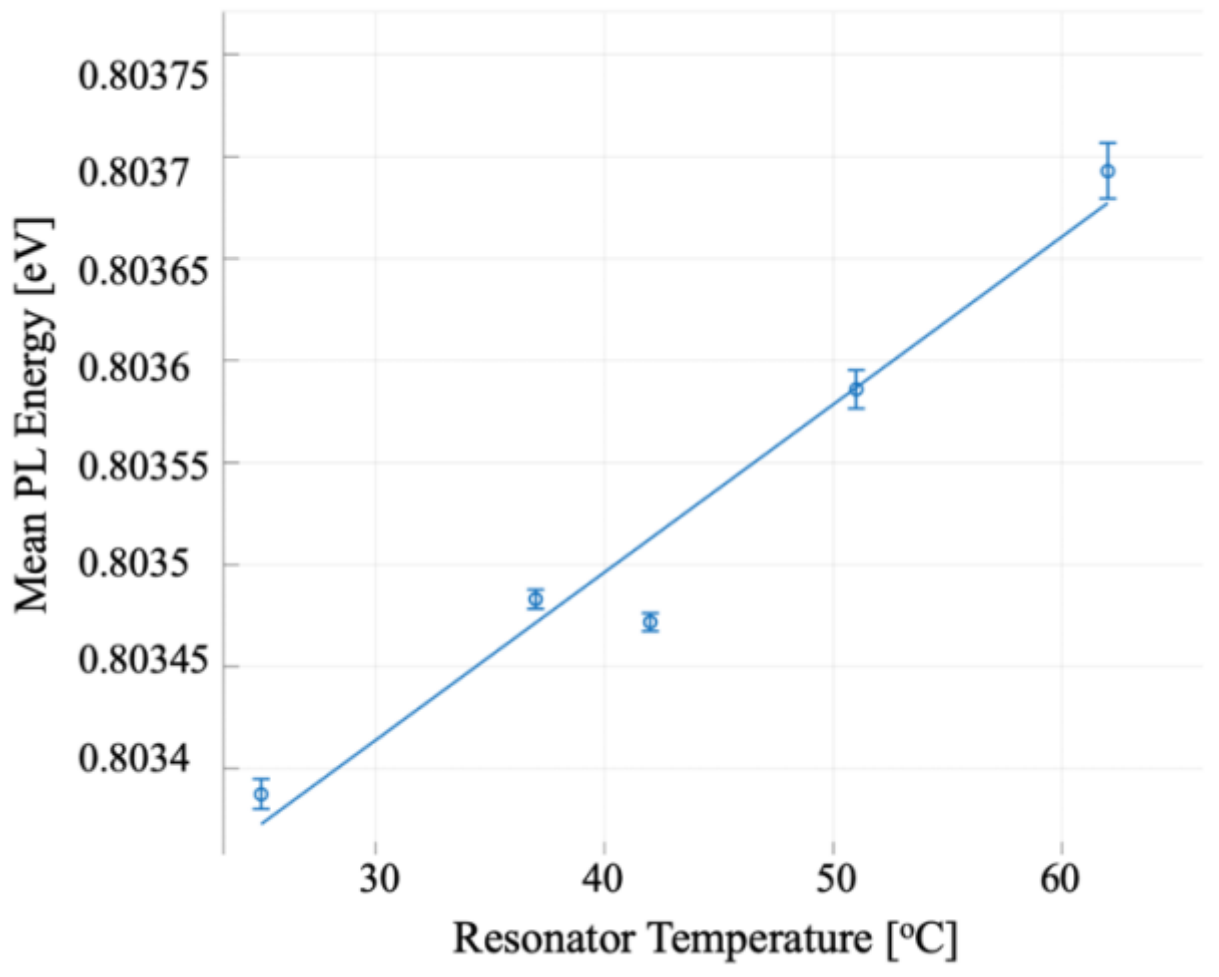

**Figure 4. Mean photoluminascence energy change with resonator temperature**. For a single pump wavelength of 1565 nm, 5 different temperatures ranging from 25 to 62ºC were taken. A standard deviation is shown for the 3 different repetitions mentioned in Fig. 1 at each temperature. linear graph with a slope of 8.23 $\mu$eV/ºC and $R^2 = 0.9589$ was fitted to the measurements.

### 3.4 Pump wavelength dependence

Finally, we vary the pump wavelength to identify where the measured spectral response is preferable. Using theoretical $Er^{3+}$ absorption and emission spectrum from [3] , we first calculate the average energy of the emission spectrum and its corresponding wavelength. Then we calculate the absorption-weighted theoretical graph based on formula 3:

$$3)\ \left(\overline{E}_{emission} - E_{pump}\right) \cdot \sigma_{absorption}$$

Figure 5(b) shows the calculated response over the full wavelength range, with a maximum at $1562.65\ nm$. Figure 5(a) compares the theoretical model with experimental measurements of the same resonator over the range of $1550 - 1570\ nm$. The blue theoretical graph is possibly a good fit for the experimental measurements which their error bars is taken from 3 repetitions. Thus, 1562.65 nm is the predicted optimum for the specified spectral figure of merit under these experimental conditions.

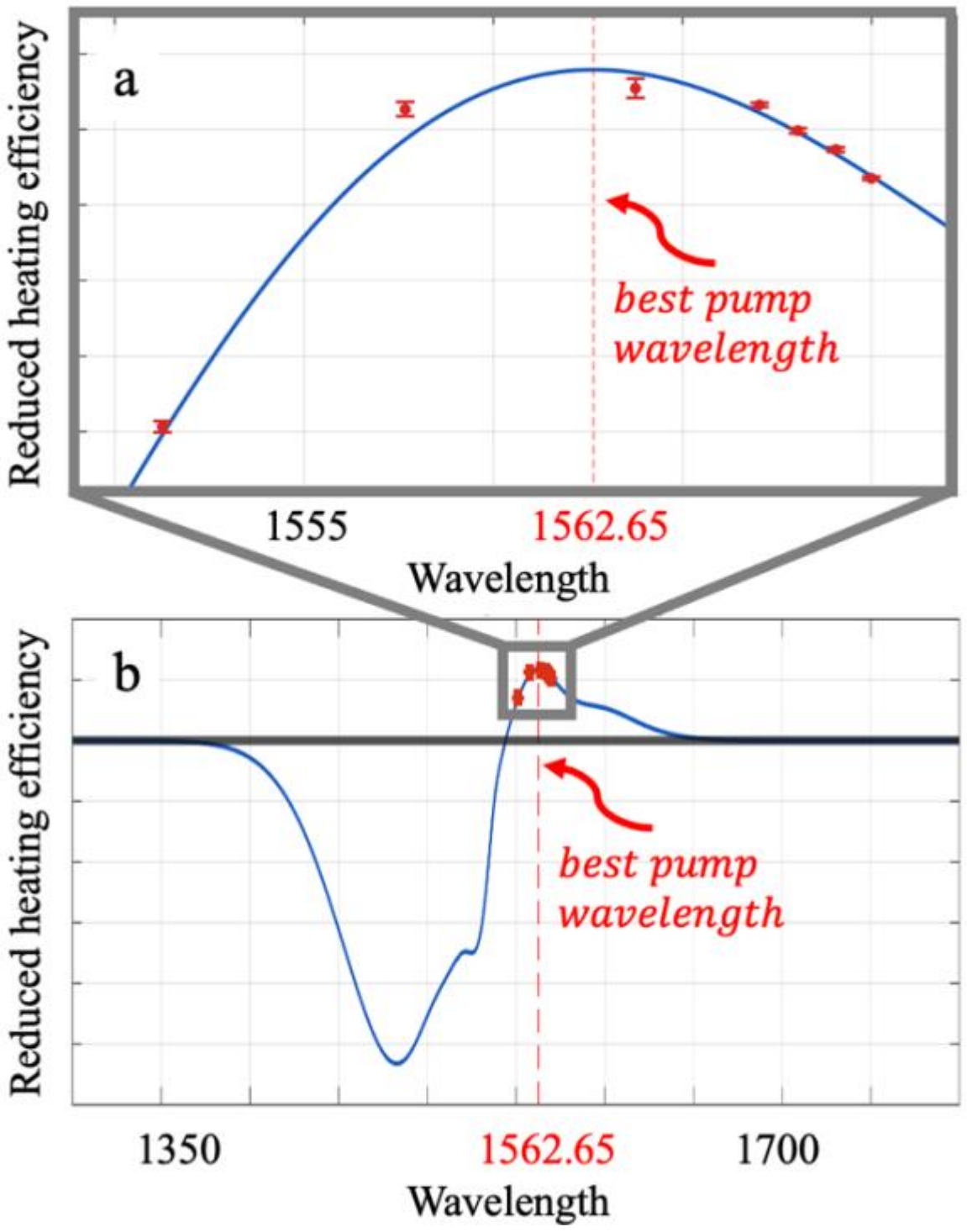

**Figure 5. Best pump for reduced heating for entire**. Theoretical model as blue graph $f(\lambda) = \sigma_{absorption}(\lambda) \cdot \left(\nu_{emission,avg} - \nu_{pump}(\lambda)\right)$ based on $Er^{+3}$ absorption spectrum[2], whereas the red dots show experimental results with their standard deviation. From theory we concur the best pump by the model prediction should be 1562.65nm. **a**. Enlargement of the region of interest. The measurements shown in red are in fact a close match to the theoretical calculation in blue. **b**. Calculation for the whole $Er^{+3}$ emission spectrum range.

## 4. Discussion

### 4.1 Interpretation and limitations

Net cooling would depend on numeral factors, such as pump power absorbed versus not-absorbed, mode volume, ration between radiative versus non-radiative transition from excited level, absorption into the silica and $Er^{+3}$ ion density in the sol-gel after annealing.

Determining the wavelength that minimizes the resonator temperature additionally requires a measurement of absorbed pump power and an independent thermal measurement.

### 4.2 Proposed dual-frequency pumping

The resonance-dependent emission patterns in Fig. 2 motivate a scheme for distributing the pump field more uniformly over the $Er^{3+}$-doped coating considering averaging over time. We propose simultaneously exciting two higher-order whispering-gallery modes at distinct wavelengths. In the calculation in Fig. 6, the polar lobes of one mode fill the gaps between those of the other near the equator, yielding a more uniform time-averaged pump-field profile across that band.

We modeled the resonator as a homogeneous silica sphere of diameter 106 µm and refractive index 1.44. The two calculated TE modes have indices $(l, m, q) = (272{,}252{,}4)$ and $(279{,}260{,}3)$ and wavelengths of 1564.91 and 1563.42 nm, respectively. We assigned the two modes equal peak electric-field amplitudes. Over times long compared with their 5.49 ps beat period, the interference term averages to zero and the squared field is the sum of the two modal contributions.

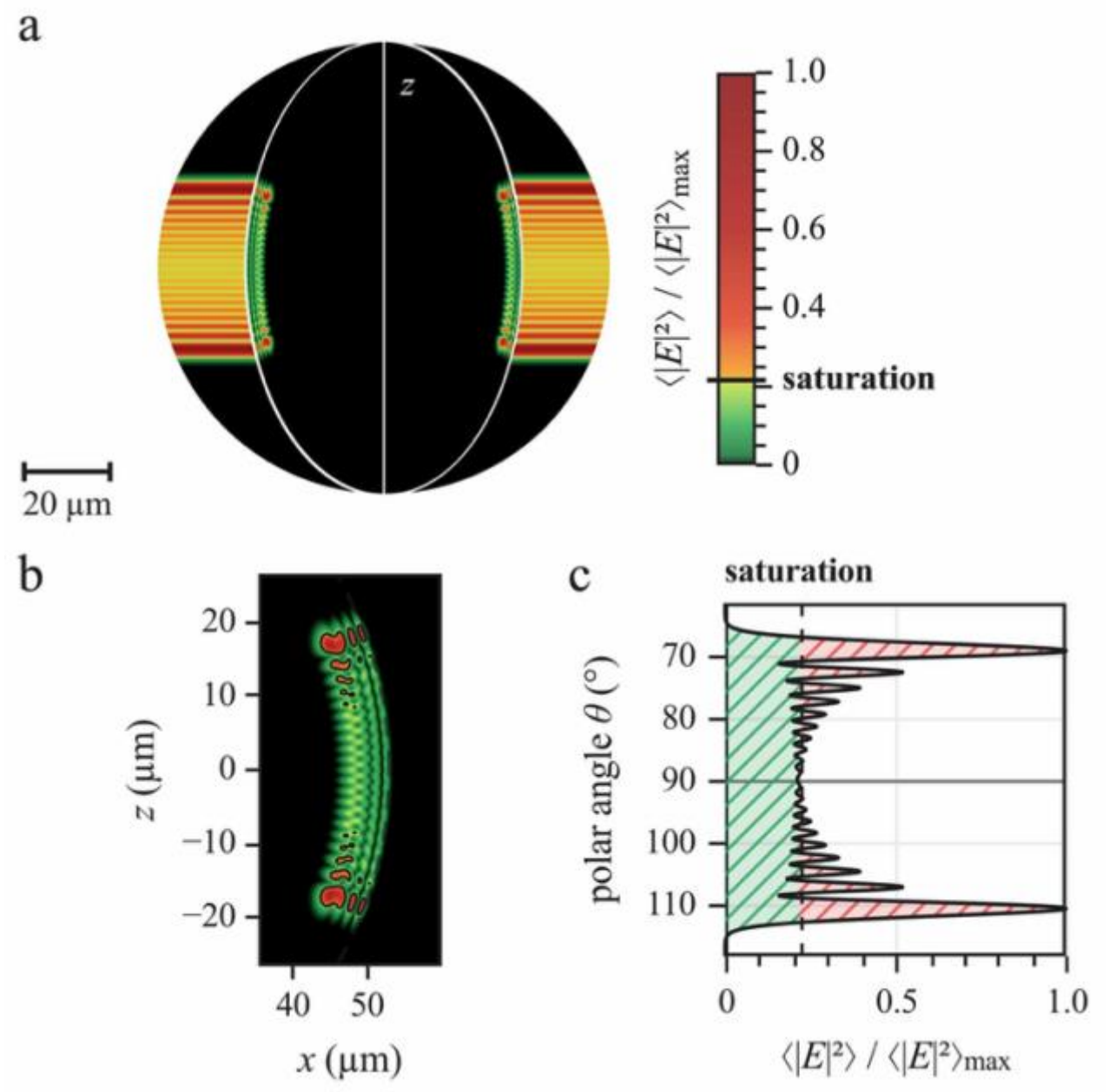

**Figure 6.** Time-averaged normalized squared electric field, ⟨|E|²⟩/⟨|E|²⟩max, for two whispering-gallery modes in a 106 µm fused-silica microsphere: A, (l, m, p, q) = (272, 252, 21, 4); B, (l, m, p, q) = (279, 260, 20, 3). (a) Equatorial view with a 75° wedge removed; scale bar, 20 µm. (b) Meridional cross-section (y = 0); the dashed curve marks the sphere surface and the black contour marks the indicated saturation threshold (0.22 of the maximum). (c) Polar profile at φ = 0°; green and red shading denote contributions below and above this threshold, respectively. The horizontal line marks the equator (θ = 90°). Only the combined field is shown.

To determine whether dual-frequency pumping reduces heating, a follow-up experiment could compare the resonator temperature under single- and dual-frequency excitation at the same total absorbed pump power.

## 5. Conclusion

Together, Figs. 1–5 show fluorescence with a mean photon energy above the pump energy, its response to controlled heating, and a pump-wavelength dependence consistent with the absorption-weighted model. The calculated profiles suggest that simultaneous excitation of two higher-order modes could distribute the pump field more uniformly near the equator. A claim of net cooling of the resonator requires measuring its temperature or dimensions under optical pumping.